\documentclass[prd,preprint,floatfix,nofootinbib]{revtex4}
\usepackage{epsfig}
\usepackage{graphicx}
\usepackage{pstricks}
\usepackage{amsmath}

\def\lsim{\mathrel {\vcenter {\baselineskip 0pt \kern 0pt
    \hbox{$<$} \kern 0pt \hbox{$\sim$} }}}
\def\gsim{\mathrel {\vcenter {\baselineskip 0pt \kern 0pt
    \hbox{$>$} \kern 0pt \hbox{$\sim$} }}}
\def\slashchar#1{\setbox0=\hbox{$#1$}           % set a box for #1
 \dimen0=\wd0                                 % and get its size
  \setbox1=\hbox{/} \dimen1=\wd1               % get size of /
\ifdim\dimen0>\dimen1                        % #1 is bigger
  \rlap{\hbox to \dimen0{\hfil/\hfil}}      % so center / in box
  #1                                        % and print #1
  \else                                        % / is bigger
 \rlap{\hbox to \dimen1{\hfil$#1$\hfil}}   % so center #1
   /                                         % and print /
  \fi}                                         %
\def\cpto{\mathrel {\vcenter {\baselineskip 0pt \kern 0pt
    \hbox{$CP$} \kern 0pt \hbox{$\longrightarrow$} }}}
\def\cptof{\mathrel {\vcenter {\baselineskip 0pt \kern 0pt
    \hbox{$~CP$} \kern 0pt \hbox{$\longleftrightarrow$} }}}

\newcommand{\ttb}{t\bar{t}}

\allowdisplaybreaks

\begin{document}

\baselineskip=15pt

\preprint{}

\title{Color-octet scalars and potentially large $CP$ violation at the LHC}

\author{Xiao-Gang He$^{1,2}$, German Valencia$^{3}$ and Hiroshi Yokoya$^{2,4}$}

\email{hexg@phys.ntu.edu.tw,valencia@iastate.edu,hyokoya@hep1.phys.ntu.edu.tw}

\affiliation{$^1$ INPAC, Department of Physics, Shanghai Jiao Tong
University, Shanghai, China \\
$^{2}$ Department of Physics, National Taiwan University, Taipei, Taiwan \\
$^{3}$ Department of Physics, Iowa State University, Ames, IA 50011 \\
$^{4}$ National Center for Theoretical Sciences, National Taiwan
University, Taipei, Taiwan }

\date{\today}

\vskip 1cm
\begin{abstract}

We consider the phenomenology of $CP$ violation in a color-octet extended scalar sector for $t\bar{t}$ production and decay at the LHC. In particular we study the effect of the two neutral color-octet scalars $S_I$ and $S_R$ that occur in the model. There are two new sources of $CP$  violation: a phase in the couplings of $S_{I,R}$ to top-quarks; and two phases in the quartic couplings of the scalar potential. In resonant production of a single $S_{I,R}$ followed by its decay into $t\bar t$ pairs through the parton level process $gg \to S_{I,R} \to t\bar t$, we find large raw $CP$ asymmetries which can reach 12\%. These raw asymmetries are, of course, diluted by standard model (SM) $t\bar{t}$ pairs making observation of $CP$ violation  contingent on whether the resonance itself can be extracted from the SM background.
\end{abstract}

\pacs{PACS numbers: 12.15.Ji, 12.15.Mm, 12.60.Cn, 13.20.Eb,
13.20.He, 14.70.Pw}

\maketitle

\section{Introduction}

Violation of charge-parity ($CP$) symmetry beyond the standard model (SM) has yet to be observed but we suspect that it must be there in order to explain
the baryon asymmetry of the universe. This gives paramount importance to
new searches for $CP$ violation in high-energy frontier. One tool, proposed many years ago, for searches in collider experiments is the use of triple-product correlations~\cite{tprods}, which are simple kinematic correlations of the form $\vec{p}_1\cdot(\vec{p}_2\times\vec{p}_3)$.
These correlations are referred to as ``naive-$T$'' odd because they  reverse sign under the ``naive-$T$'' operation that reverses the direction of momenta and spin without interchanging initial and final states. In general, correlations of this form can be $CP$ even or $CP$ odd, as they are induced by either loop level unitarity phases or by $CP$ violating phases respectively. In top-quark pair production, these triple-product correlations originate in $CP$ violating spin correlations with  the top-quark (and anti-quark) weak decay acting as spin analyzer.

There exist several recent proposals to search for $CP$ violation  at the LHC using triple-product correlations. Our discussion is based on observables discussed for anomalous top-quark couplings in Ref.~\cite{cplhc} as well as observables discussed for multi-Higgs models in Ref.~\cite{cphiggs}. Additional processes that have been discussed recently include $W$ and $Z$ pair production and decay~\cite{others}.
On the other hand, $CP$ conserving ``naive-$T$'' odd triple-product correlations have been studied
in radiative top-quark decay~\cite{Hagiwara:2007sz} at one-loop level.

In this paper we consider triple-product correlations in top-quark pair production at the LHC, induced by new $CP$ violating interactions in a scalar sector extended with a color-octet electroweak-doublet as described in Ref.~\cite{Manohar:2006ga}. This model incorporates the additional scalars in a manner consistent  with minimal flavor violation (MFV) in order to naturally suppress flavor changing neutral currents (FCNC).

Of particular interest to us are the two neutral, color-octet, scalar
resonances that occur in the model, $S_{I,R}$. These particles couple at
the one-loop level to two gluons and their production at the LHC has
been discussed in Ref.~\cite{Gresham:2007ri} for the $CP$ conserving
case. These particles also couple (dominantly) to top-quark pairs which
makes them an ideal candidate to study $CP$ violation in the process $gg
\to S_{I,R} \to t \bar{t}$. To this end, we extend the results  of
Ref.~\cite{Gresham:2007ri} to include $CP$ violation, which can occur
both in the couplings of $S_{I,R}$ to top-quark pairs as well as in
certain self-interactions in the scalar potential. We further assume
that top-quark decay proceeds as in the SM and serves only to analyze the corresponding spin. Within this framework we find that relatively large raw asymmetries are possible, as large as $\sim12\%$. These raw asymmetries are diluted by the SM  top-quark pairs and their observation is contingent on the resonance itself being observable. For illustration, we present a set of parameters for which the resonance is visible over the SM background and the resulting $CP$ asymmetry can be as large as a few percent. We also comment on other channels where $CP$ asymmetries are potentially visible in cases where the  single resonance is not.

The LHC has already established new constraints on the new physics
beyond the SM, even though it has been operating so far with a reduced
energy of 7~TeV. In particular, both ATLAS and CMS have excluded certain
color-octet scalars similar to the ones we consider here in a broad mass
range~\cite{noreson} by studying the dijet channel. These exclusion
limits, while interesting, do not apply to the models we discuss, where
the color-octet resonances decay almost exclusively into top-quark
pairs. Their decay modes into dijets occur with branching ratios below
$10^{-3}$ for the sets of parameters we use in this study.

Our paper is organized as follows: in Section II we review the relevant aspects of the model for the color-octet scalars with emphasis on the $CP$ violating phases that have not been studied previously. In Section III we study several benchmark cases that illustrate the generic properties of the raw $CP$ asymmetries. We also discuss several aspects concerning the observability of the $CP$ violating signals at the LHC. In Section IV we state our conclusions and, finally,  we relegate some analytic formulae to the Appendix.

\section{Color-octet scalars as a source for $CP$ violation}

We briefly review the case of a scalar sector that has been extended
with a color-octet electroweak-doublet scalar with hypercharge $1/2$, $O=(8,2,1/2)$. This particular choice is
motivated by the requirement of MFV and has been recently elaborated in Ref.~\cite{Manohar:2006ga,Gresham:2007ri}. It
was noted in these papers that in a MFV scenario only scalars with the same gauge quantum numbers as the SM Higgs doublet $H = (1,2,1/2)$
or color-octet scalars with the same weak quantum number as the Higgs
doublet $O = (8,2,1/2)$ can couple to quarks, and this has many
interesting consequences for both collider and flavor physics. This
color-octet electroweak doublet can be written in the
properly normalized component form with the color index $A$ as $O = \sqrt{2} S =\sqrt{2}\ T^A (S^{A+}, S^{A0})^T$, where $T^A$ is the $SU(3)_C$ generator normalized as ${\rm Tr}(T^AT^B) = \delta^{AB}/2$.

The Yukawa couplings of the color-octet scalars can be parameterized, to
the leading order, with the MFV assumption as~\cite{Manohar:2006ga}
\begin{eqnarray}
{\cal L} &=& - {\sqrt{2}\over v} \tilde \eta_{U} \bar U_R T^A \hat M^u U_LS^{A0} + {\sqrt{2}\over v} \tilde \eta_{U} \bar U_R T^A \hat M^u V_{KM} D_L S^{A+}\nonumber\\
&-& {\sqrt{2}\over v} \tilde \eta_{D} \bar D_R T^A \hat M^d D_LS^{A0\dagger} - {\sqrt{2}\over v} \tilde \eta_{D} \bar D_R T^A \hat M^u V^\dagger_{KM} U_L S^{A-} + h.c.\,,
\end{eqnarray}
where $\hat{M}^{u,d}$ are the diagonalized mass matrices, $\hat{M}^{u,d} = {\rm diag}(m_{u,d}, m_{c,s}, m_{t,b})$; $U_{L,R}$ and $D_{L,R}$ are the up and down quarks,
$U_{L,R} = {\rm diag}(u_{L,R}, c_{L,R}, t_{L,R})$ and $D_{L,R} = {\rm
diag}(d_{L,R}, s_{L,R}, b_{L,B})$; and $v\sim 246$~GeV is the Higgs
vacuum expectation value (VEV), $\langle H \rangle = v/\sqrt{2}$. The neutral complex field $S^{A0}$ can be further decomposed into a scalar $S^{A0}_R$ and a pseudo-scalar $S^{A0}_I$ as $S^{A0} = (S^{A0}_R + iS^{A0}_I)/\sqrt{2}$.
The parameters $\tilde \eta_{U,D}$ are expected to be of order one and are in general complex. We will write them as $\tilde \eta_{U,D} = \eta_{U,D}e^{i\alpha_{u,d}}$ with $\eta_{U,D}$ real, and if there are
non-zero phases $\alpha_{u,d}$ there is $CP$ violation beyond the SM.

There is a second possible source of $CP$ violation in this model in two of the self couplings appearing in the scalar potential. The most general potential with the complex scalar doublets $H$ and $S$ is given by~\cite{Manohar:2006ga},
\begin{eqnarray}
V &=& {\lambda\over 4} \left ( H^{\dagger i} H_i - {v^2\over 2}\right )^2 + 2 m^2_s {\rm Tr}\ S^{\dagger i}S_i + \lambda_1 H^{\dagger i} H_i {\rm Tr}\ S^{\dagger j}S_j + \lambda_2 H^{\dagger i}H_j
{\rm Tr}\  S^{\dagger j}S_i \nonumber\\
&+& [\tilde \lambda_3 H^{\dagger i} H^{\dagger j} {\rm Tr}\ S_i S_j + \tilde \lambda_4 H^{\dagger i}{\rm Tr}\ S^{\dagger j}S_j S_i + \tilde \lambda_5 H^{\dagger i}{\rm Tr}\ S^{\dagger j}S_i S_j + h.c.]\nonumber\\
&+&\lambda_6 {\rm Tr}\  S^{\dagger i} S_i S^{\dagger j}S_j + \lambda_7 {\rm Tr}\  S^{\dagger i} S_j S^{\dagger j}S_i + \lambda_8 {\rm Tr}\  S^{\dagger i} S_i{\rm Tr}\  S^{\dagger j}S_j \nonumber\\
&+&\lambda_9 {\rm Tr}\  S^{\dagger i} S_j {\rm Tr}\  S^{\dagger j}S_i + \lambda_{10} {\rm Tr}\  S_i S_j S^{\dagger i}S^{\dagger j} + \lambda_{11} {\rm Tr}\  S_i S_j S^{\dagger j}S^{\dagger i}\;.
\end{eqnarray}
The parameters $\tilde \lambda_{3,4,5}$ are in general complex, but without loss of generality, one can  choose a convention in which $\tilde \lambda_3=\lambda_3$ is real. The two other phases cannot be removed and we write them as $\tilde \lambda_{4,5}= \lambda_{4,5} e^{i\alpha_{4,5}}$. Non-zero phases $\alpha_{4,5}$ provide the second source of $CP$ violation beyond the SM present in the model.
We note the custodial symmetry requires
$\lambda_4=\lambda_5^*$~\cite{Burgess:2009wm,Carpenter:2011yj}, so that 
$\alpha_4=-\alpha_5$ in our parameterization.

If the mass of color-octet scalars is not too large, a tree-level interaction can pair produce them at the LHC through the process $gg \to S^\dagger S$. Since the color-octet scalars also couple to quarks, there is an additional contribution from $q \bar q \to S^\dagger S$. However this contribution is small because the Yukawa couplings of $S$ to quarks are proportional to quark masses. Single $S$ production is also possible at tree level from its Yukawa couplings to quarks, but this tree-level contribution is small as it is proportional to the light-quark mass. It has been shown that
the single $S$ production cross section at the LHC is dominated by a
loop induced $gg-S$ interaction and can be of order $100$~fb for masses
of order a few hundred GeV to a TeV~\cite{Gresham:2007ri}. If these
resonances are produced at the LHC, it will be possible to study their
properties, including $CP$ violation, through their decays to SM
particles. The dominant decay mode is $S\to t\bar t$ although there is
also one-loop induced decay into gluon pair: $S\to
gg$~\cite{Manohar:2006ga,Gresham:2007ri}.

In this paper we wish to study signals of $CP$ violation at the
LHC. This is accomplished by first observing the resonant production of
the neutral color-octet scalars $S_{I,R}$ through their effective $gg-S$
couplings and then studying the correlations that occur in the
subsequent decay chain: $gg \to S_{I,R} \to t \bar t \to b \bar{b} \mu^+
\mu^- \nu \bar \nu$. The new sources of $CP$ violation induce
triple-product correlations involving the spin of the top and anti-top
quarks which are subsequently analyzed by their weak decays. In the
dimuon mode chosen above, it is the direction of the muons that analyzes
the spin directions. Both the new sources of $CP$ violation contribute
to the resulting asymmetries.

\begin{figure}%[htb]
\centerline{
\includegraphics[angle=0, width=.85\textwidth]{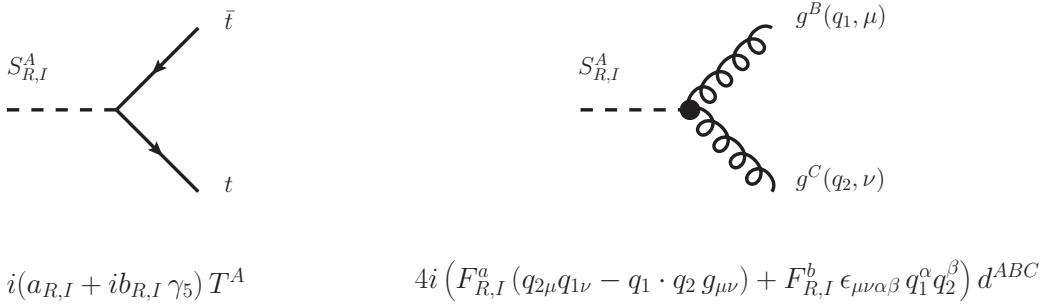}%{feynrules}
}
\caption{\small\sf Feynman Rules.}
\label{f:feyndiags}
\end{figure}

The relevant  effective couplings are shown schematically in Figure~\ref{f:feyndiags}, and they can be written in terms of an effective Lagrangian as
\begin{eqnarray}
{\cal L}(S-t\bar t) &=& \bar t (a_R + i b_R\gamma_5) T^A  t S^{A0}_R + \bar t (a_I + i b_I\gamma_5) T^A  t S^{A0}_I\;,\nonumber\\
{\cal L}(S-g g )&=& (F^a_R G^A_{\mu\nu} G^{B\mu\nu} + F^b_R \tilde G^A_{\mu\nu}G^{B\mu\nu})d^{ABC}S^{A0}_R\nonumber\\
&+& (F^a_I G^A_{\mu\nu} G^{B\mu\nu} + F^b_I \tilde G^A_{\mu\nu}G^{B\mu\nu})d^{ABC}S^{A0}_I\;,
\end{eqnarray}
where $G^A_{\mu\nu}$ is the gluon field strength tensor and $\tilde G^{A\mu\nu} = (1/2)\epsilon^{\mu\nu\alpha\beta}G^A_{\alpha\beta}$.

%--------------------------------------------------------------------
%--------------------------------------------------------------------
The couplings to the top-quark occur at tree-level and they contain $CP$ violation originating in the new phase  $\tilde{\eta}_U$. They are given by
\begin{eqnarray}
&&S^{A0}_R - t\bar t\;:\;\; i(a_R + ib_R\gamma_5) = -i\eta_U{m_t\over v}(\cos\alpha_u - i \sin\alpha_u \gamma_5)T^A\;,\nonumber\\
&&S^{A0}_I - t\bar t\;:\;\; i(a_I + i b_I\gamma_5) = i\eta_U{m_t\over v}(\sin\alpha_u + i \cos \alpha_u \gamma_5)T^A\;.
\end{eqnarray}
The couplings to gluons occur at one-loop level and can be easily derived following Ref.~\cite{Gresham:2007ri}. The $CP$ violation in this case is due to phases in the couplings $\tilde{\lambda}_{4,5}$. We find
\begin{eqnarray}
F^a_R &=& (\sqrt{2}G_F)^{1/2} {\alpha_s\over 8\pi}\left[ \eta_U \cos\alpha_u\, I_q\left(\frac{m^2_t }{m_R^2}\right) \right. \nonumber \\
&-& \left. {9\over 4} {v^2\over m^2_R} (\lambda_4 \cos\alpha_4 + \lambda_5 \cos\alpha_5) \left\{\frac{1}{2}I_s(1)+\frac{1}{2}I_s\left(\frac{m_I^2}{m_R^2}\right)\right\}\right]\;,\nonumber\\
F^b_R &=& (\sqrt{2}G_F)^{1/2} {\alpha_s\over 8\pi}{1\over 2}{m^2_t \over m^2_R} \eta_U \sin\alpha_u\, f\left(\frac{m^2_t}{m_R^2}\right),\nonumber\\
F^a_I &=& (\sqrt{2}G_F)^{1/2} {\alpha_s\over 8\pi}\left[ - \eta_U  \sin\alpha_u\,I_q\left(\frac{m^2_t }{m_I^2}\right) \right. \nonumber \\
&+& \left. {9\over 4} {v^2\over m^2_I} (\lambda_4 \sin\alpha_4 + \lambda_5 \sin\alpha_5) \left\{\frac{5}{6}I_s(1)+\frac{1}{6}I_s\left(\frac{m_R^2}{m_I^2}\right)\right\}\right]\;,\nonumber\\
F^b_I &=& (\sqrt{2}G_F)^{1/2} {\alpha_s\over 8\pi}{1\over 2} {m^2_t
 \over m^2_I} \eta_U \cos\alpha_u\, f\left(\frac{m^2_t}{m_I^2}\right).
\end{eqnarray}
In these expressions we have assumed that the mass of the charged color-octet scalars $S^\pm$ is equal to the mass of $S_I$, $m^\pm = m_I$, which corresponds to the custodial symmetry conserving case where $2\lambda_3=\lambda_2$~\cite{Manohar:2006ga}. We have allowed for the mass of $S_R$ to be different, $m_R\neq m_I$. These two masses are related by $m_R^2-m_I^2 =\lambda_3 v^2$. Throughout the calculation, the scalars $S_R$ and $S_I$ (when not in a loop), are taken to be on-shell, in keeping with the narrow width approximation.
The loop functions $I_{q,s}$ and $f$ are defined by:
\begin{eqnarray}
I_q(z) &=& 2z+z(4z-1)f(z), \quad %\nonumber \\
I_s(z) = -z(1+2zf(z)), \nonumber \\
f(z) &=& \frac{1}{2}\left(\ln\left(\frac{1+\sqrt{1-4z}}{1-\sqrt{1-4z}}\right)-i\pi\right)^2 \ \ {\rm for~}z<1/4 \nonumber \\
&=&  -2 \left(\arcsin\left(\frac{1}{2\sqrt{z}}\right)\right)^2 \ \ {\rm for~}z>1/4 .
\label{loopintegrals}
\end{eqnarray}
From this it follows that $I_s(1)=\pi^2/9-1$, a factor that generates some suppression in contributions from scalar loops relative to top-quark loops.

\section{Estimate of CP-odd asymmetries}

We will now give numerical estimates for the triple-product correlations that will serve as $CP$-odd observables. Following Ref.~\cite{cplhc}, we know that the best observable for the case of $t\bar{t}$ production and decay is the correlation
\begin{eqnarray}
 \tilde {\cal{O}}_1 &=&
  \epsilon_{\mu\nu\alpha\beta}\,p_b^\mu\,p_{\bar{b}}^\nu\,
  p_{\mu^+}^\alpha\,p_{\mu^-}^\beta \,\,
  \xrightarrow[]{b\bar b ~CM}\,\, \propto \,\, \vec{p}_b\cdot
  (\vec{p}_{\mu^+} \times \vec{p}_{\mu^-}).
  \label{cpcorrelation}
\end{eqnarray}
The first, covariant, expression is given in terms of the completely
antisymmetric Levi-Civita tensor, whereas the second one indicates its
reduction to a simple triple-product correlation in the $b\bar{b}$
center-of-mass frame. The reasons to select this correlation from the
set described in Ref.~\cite{cplhc} are twofold. First, the dimuon decay
of $t\bar{t}$ is the cleanest. Second, the fact that the $t\bar{t}$ pair
is produced from a scalar intermediate state prevents the appearance of
correlations involving the beam momentum. Note that although it appears
that this correlation requires distinguishing the $b$ and $\bar{b}$
jets, it is only necessary to systematically associate one of the $b$
jets with one of the muons. For example, the "$b$"-jet could be the one
closest to the $\mu^+$.

To study the effect of the correlation  Eq.~(\ref{cpcorrelation}) we
consider the laboratory frame distribution $d\sigma/d{\tilde{\cal O}}_1$.
The $CP$ violating effects can be isolated by extracting asymmetric terms from this distribution, either by a direct fit or by 
constructing the integrated counting
asymmetry
\begin{eqnarray}
 A_1 &\equiv & \frac{N_{events}(\tilde{\cal O}_1
  >0)-N_{events}(\tilde{\cal O}_1<0)}{N_{events}(\tilde{\cal
  O}_1>0)+N_{events}(\tilde{\cal O}_1<0)}.
  \label{asym}
\end{eqnarray}

To measure this distribution (and its associated integrated asymmetry)
we generate events for the process $pp\to S_{I,R} \to t{\bar t} \to b \mu^+ \nu_\mu
{\bar b} \mu^-\bar{\nu}_\mu$ with the aid of {\tt
MadGraph}~\cite{madgraph}. To generate the signal events we implement
the vertices of Figure~\ref{f:feyndiags} into the {\tt MadGraph} code. We also use the default {\tt MadGraph} SM processes to generate the corresponding events. 
In all cases we use the default {\tt MadGraph} cuts requiring
the top quark and $W$ boson intermediate states to be within 15 widths
of their mass shell, the transverse momentum $p_T$ of both muons to be
larger than 10~GeV and the pseudo-rapidity of both muons to be
$|\eta_\mu| < 2.5$.
We also use SM parameter values as in {\tt MadGraph} and the CTEQ-6L1
parton distribution functions~\cite{Pumplin:2002vw}. \\

\begin{table}[t]
 \begin{center}
  \begin{tabular}{|c||c|c|c|c|}
   \hline
   & \shortstack{Parameters\\{}}
   & \shortstack{Decay Width\\{}[GeV]}
   & \shortstack{Resonance\\[1mm] cross-section [fb]}
   & \shortstack{\\[1mm]Raw asymmetry\\around resonance} \\
   \hline \hline
   \shortstack{Case~1\\{}} &
   \shortstack{\\[1mm] $m_R,m_I=500,700$~GeV,\\
   $\eta_U=\lambda_{4,5}=1$, $\alpha_u=\pi/4$} &
   \shortstack{$S_R$ : 2.7\\ $S_I$ : 5.3} &
   \shortstack{$S_R$ : 5.9\\ $S_I$ : 2.8} &
   \shortstack{$S_R$ : $A_1=-0.123$ \\ $S_I$ : $A_1=0.121$}
   \\[1mm] \hline
   \shortstack{Case~2\\{}} &
   \shortstack{\\[1mm] $m_R,m_I=500,700$~GeV,\\
   $\eta_U=1, \lambda_{4,5}=8$, $\alpha_{u,4}=-\alpha_5=\pi/4$} &
   \shortstack{$S_R$ : 2.7 \\ $S_I$ : 5.3} &
   \shortstack{$S_R$ : 6.5\\ $S_I$ : 2.8} &
   \shortstack{$S_R$ : $A_1=-0.125$ \\ $S_I$ : $A_1=0.123$}
   \\[1mm] \hline
   \shortstack{Case~3\\{}} &
   \shortstack{\\[1mm] $m_R,m_I=500,700$~GeV,\\
   $\eta_U=\lambda_{4,5}=1$, $\alpha_{u}=\pi/8$} &
   \shortstack{$S_R$ : 2.1 \\ $S_I$ : 5.8} &
   \shortstack{$S_R$ : 7.2\\ $S_I$ : 3.1} &
   \shortstack{$S_R$ : $A_1=-0.112$ \\ $S_I$ : $A_1=0.084$}
   \\[1mm] \hline
   \shortstack{Case~4\\{}} &
   \shortstack{\\[1mm] $m_R=m_I=500$~GeV,\\
   $\eta_U=\lambda_{4,5}=1$, $\alpha_{u}=\pi/8$} &
   \shortstack{$S_R$ : 2.1 \\ $S_I$ : 3.3} &
   \shortstack{$S_R+S_I$ : \\ \quad  12.3} &
   \shortstack{$S_R+S_I$ : \\ \quad $A_1=-0.017$}
   \\[1mm] \hline
   \shortstack{Case~5\\{}} &
   \shortstack{\\[1mm] $m_R,m_I=400,1000$~GeV,\\
   $\eta_U=\lambda_{4,5}=1$, $\alpha_{u}=\pi/4$} &
   \shortstack{$S_R$ : 1.2 \\ $S_I$ : 8.8} &
   \shortstack{$S_R$ : 8.9\\ $S_I$ : 1.1} &
   \shortstack{$S_R$ : $A_1=-0.089$ \\ $S_I$ : $A_1=0.096$}
   \\[1mm] \hline
  \end{tabular}
 \end{center}\caption{Parameter values, resonance $S_{I,R}$ decay-widths and production cross-sections $\sigma(pp\to S_{I,R}\to t \bar{t} \to b \bar{b} \mu^+\mu^- \nu\bar{\nu})$ at the LHC $\sqrt{S}=14$~TeV, and  raw $CP$ asymmetry for  the five cases discussed in the text. 
 The raw asymmetry is defined by taking into account the events with
 $|m_{\ttb}-m_{I,R}|<10$~GeV.}\label{tab1}
\end{table}

We begin by presenting raw asymmetries (no SM top-quark pair events) using values for the parameters $\eta_U$ and $\lambda_{4,5}$ in the ranges discussed in Ref.~\cite{Gresham:2007ri}, as well as $m_{I,R} \lsim 1$~TeV. We illustrate the $CP$ asymmetries utilizing only the dimuon signal, but emphasize that other decay modes can also be used.

The results are summarized in Table~\ref{tab1} where we show for each set of parameters the corresponding resonance ($S_{I,R}$) widths (which are completely dominated by the $t\bar{t}$ decay mode); production cross-sections at the LHC $\sigma(pp\to S_{I,R}\to t \bar{t} \to b \bar{b} \mu^+\mu^- \nu\bar{\nu})$ for $\sqrt{S}=14$~TeV; and a raw asymmetry. This raw asymmetry is estimated separately for each resonance by including only top-quark pair events that result from the decay of that resonance and have an invariant mass within 10~GeV of the resonance,  $|m_{t\bar{t}}-m_{I,R}|<10$~GeV.

\begin{figure}[t]
\centerline{
\includegraphics[angle=0, width=.65\textwidth]{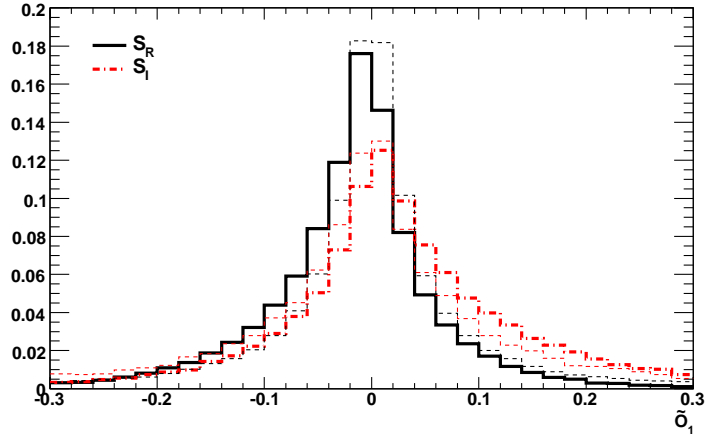}
}
\caption{$d\sigma/d\tilde{O}_1$ distributions for the events with
 $|m_{\ttb}-m_{R,I}|<10$~GeV in Case~1, where the SM contribution is
 omitted.
 Thin dashed lines are for the zero $CP$ phase case, but the same masses
 and couplings as in Case~1.}\label{fig:o1}
\end{figure}

We begin with Case 1 in which we have  two well separated resonances, $m_R =500$~GeV and $m_I= 700$~GeV, $|m_R-m_I| >> \Gamma_{I,R}$. The scalar potential parameters $\lambda_{4,5}$, as well as the parameter $\eta_U$ governing the strength of the $St\bar{t}$ coupling, are all chosen to be one for illustration. 
We also choose $CP$ violating phases to be
$\alpha_u=\pi/4$, $\alpha_{4,5}=0$, in order to 
maximize the $CP$ violating interaction in the coupling of scalars
to $t\bar{t}$. As seen in Table~\ref{tab1}, the resulting raw asymmetry can be rather large, about $12\%$ for each resonance. The results also show that the two resonances produce asymmetries that tend
to cancel each other out, as can be inferred from Eq.~(\ref{dianddr}). 
In Fig.~\ref{fig:o1}, we plot the $d\sigma/d\tilde{\cal O}_1$
distributions for the two scalars, where $\tilde{\cal O}_1$ is evaluated
as $\tilde{\cal O}_1 = \vec{p}_b\cdot (\vec{p}_{\mu^+} \times
\vec{p}_{\mu^-})/m_t^3$ in the $b\bar{b}$ rest-frame.  The solid and
dot-dashed lines show the $CP$ violating case whereas the thin dashed lines illustrate the corresponding $CP$ conserving case where the phases have been set to zero.

In Case 2 we choose much larger values for the parameters $\lambda_{4,5}$, taking them to be 8 times larger than $\eta_U$. With this choice we want to enhance the relative contribution from the scalar loops which are otherwise suppressed by the factor $(\pi^2/9-1)$ mentioned before. We also introduce non-zero $CP$ phases $\alpha_{4,5}$. We find that this choice produces a modest increase in the cross-section for $S_R$ but not for $S_I$. The raw $CP$ asymmetries remain  about the same indicating that the top quark loop is still dominant.

In Case 3 we repeat the parameters of Case 1 except for choosing $\alpha_U =\pi/8$ which has the effect of minimizing the cancellation of asymmetries between the two resonances.  This can be seen from the Table~\ref{tab1} where the asymmetry around $S_R$ remains near 12\% but the asymmetry near $S_I$ is down to about 8\%.  The production cross-section for each resonance is also larger with this choice of phase and the widths are affected as well with $S_R$ becoming narrower and $S_I$ wider.

As Eq.~(\ref{ffnearres}) indicates, when the two masses $m_{I,R}$ are
close to each other, the contribution from the top-quark loop tends to
cancel out, exposing the effect of $\lambda_{4,5}$. We illustrate this
in Case 4 with $m_I=m_R$ and $CP$ phase $\alpha_u=\pi/8$. The resulting
raw asymmetry is  smaller but non-vanishing.

Finally, in Case 5, we illustrate the effect of the resonance mass by
choosing one of the resonances to be just above the $t\bar{t}$ threshold
and the other one near the high-end of the LHC reach.  With the same $CP$ phases as in Case 1, the asymmetries are about 25\% smaller in this case.

We now turn to the question of observability of the raw asymmetries over
the SM background of top-quark pairs.  First, there is a matter of
statistical sensitivity.  At the LHC, the total cross-section for
$\ttb$ events is approximately $\sigma_{\ttb}=850$~pb. This implies that
the expected number of events in the dimuon channel is about $10^5$ for
an  integrated luminosity ${\cal L}=10$~[fb$^{-1}$].
Therefore, an optimistic sensitivity for the asymmetry {\it using only
the dimuon channel} for one year of nominal LHC running, is $\delta{A}_1=1/\sqrt{\sigma_{\ttb}*{\it B}_{\mu\mu}*{\cal L}}\sim
3\times 10^{-3}$, where ${\it B}_{\mu\mu}\simeq 1/81$ is the branching
ratio of the dimuon channel. Of course this can be improved
significantly by considering other decay channels. For example, in
Ref.~\cite{cplhc} it is shown how to use the lepton-plus-jets channel and purely hadronic channels to measure $CP$ asymmetries. A detailed analysis of all channels is beyond the scope of the present paper where we simply seek to establish the possibility of a large raw asymmetry. 

\begin{figure}[t]
\centerline{
\includegraphics[angle=0, width=.5\textwidth]{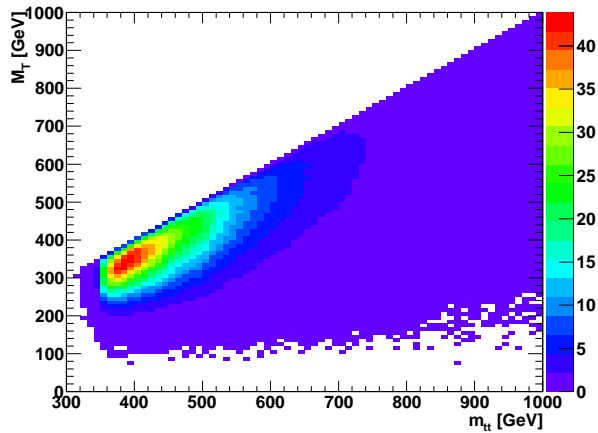}
}
\caption{Two-dimensional density plot of $m_{\ttb}$ and $M_{T}$,
 for the $\ttb$ production in the dimuon mode.}
\label{fig:MT}
\end{figure}

Second, as we have seen, the asymmetries due to the two resonances tend
to cancel. This indicates the need to isolate a certain window in
top-quark pair invariant mass around the resonance to extract a non-zero
asymmetry. In the dimuon channel it is not possible to reconstruct
$m_{t\bar{t}}$ so we use the transverse mass $M_{T}$ for this
purpose.\footnote{%
The transverse mass $M_T$ is defined as 
$M_T = \sqrt{
 \left(E\hspace{-6pt}/_T + E_{T}^{\ell\ell} + E_{T}^{b\bar{b}}\right)^2
 - \left(\vec{p\hspace{-5.5pt}/}_T + \vec{p}_{T}^{\,\ell\ell}
 + \vec{p}_{T}^{\,b\bar{b}}\right)^2}$,
where $E\hspace{-6pt}/_T=\left|\vec{p\hspace{-5.5pt}/}_T\right|$, 
$\vec{p\hspace{-5.5pt}/}_T$ is missing transverse momentum,
$\vec{p}_{T}^{\,\ell\ell (b\bar{b})}$ is vector-sum of dimuon
($b$$\bar{b}$ pair) transverse momentum, $E_T^{\ell\ell (b\bar{b})} =
\sqrt{m^2_{\ell\ell (b\bar{b})}+\left|\vec{p}_{T}^{\,\ell\ell
(b\bar{b})}\right|^2}$ with $m_{\ell\ell (b\bar{b})}$ the
invariant-mass of dimuon ($b\bar{b}$ pair).}
The correlation between these two variables is shown in
Figure~\ref{fig:MT}.

This complication would also be treated differently in other top-quark decay channels, and its resolution will ultimately depend on an observation of a new resonance in $t\bar{t}$ events. At the same time, it shows that if one wants to check for $CP$ asymmetries in $t\bar{t}$ events in which a resonance has not been observed, there are good reasons to study separately different small windows in $m_{t\bar{t}}$ (or the corresponding observable such as $M_T$), as small as allowed by statistics.

\begin{figure}[th]
\centerline{
\includegraphics[angle=0, width=.55\textwidth]{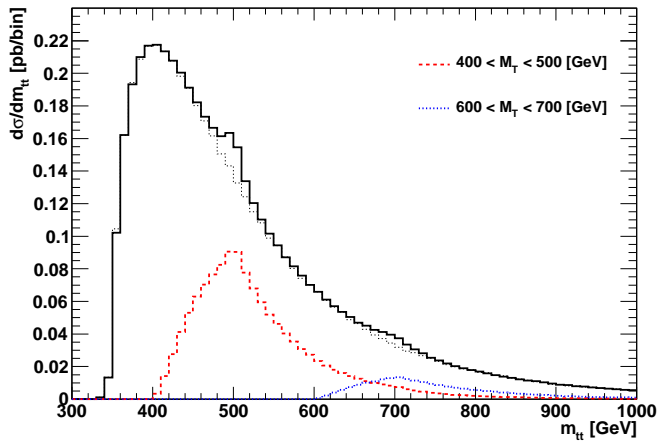}
}
 \caption{$d\sigma/dm_{\ttb}$ distributions without and with contribution
 of scalars in Case~1$^\prime$, plotted in dotted and solid line,
 respectively, in an unit of pb per 5 GeV bin.
 Dashed lines are for the distribution after $400<M_{T}<500$~[GeV] (red
 long-dashed) and $600<M_{T}<700$~[GeV] (blue short-dashed) cuts.}\label{fig:mtt}
\end{figure}

The resonant cross-section for all the cases discussed in
Table~\ref{tab1} is significantly smaller than the cross-section for SM
top-quark pairs in a 20~GeV window in $m_{t\bar{t}}$ which is $\sim 280$~fb
when the window is centered at $m_{\ttb}=500$~GeV. This number is valid at leading
order, including branching ratios for decay into dimuon channel, and
including all the kinematic cuts discussed before. The SM events will
thus dilute the raw asymmetries by two orders of magnitude making them
unobservable for practical purposes. In other words, the large raw
asymmetries in Table~\ref{tab1} are not observable at the LHC because the corresponding resonances do not stand out above the SM background. To continue our discussion we thus consider resonances with larger production cross-section.  This can be easily achieved by increasing $\eta_U=1$ to $\eta_U=3$, which results in an order of magnitude increase in the resonance cross-sections without changing the asymmetries significantly.\footnote{The value of $\eta_U=3$ in conjunction with masses below one TeV is slightly outside the parameter space allowed by $R_b$ studied in Ref.~\cite{Gresham:2007ri}. We use it anyway as a simple way to illustrate a resonance that stands above SM background taking into consideration the imprecise nature of indirect constraints on new physics.}  We illustrate this in Table~\ref{tab1'}. Notice that we have kept the definition of the raw asymmetry with a 20~GeV window around the resonance as in Table~\ref{tab1}. This window does not cover a full width in all cases with $\eta_U=3$, resulting in fractionally smaller asymmetries than would be possible. The precise optimization of the window size, although eventually important for measurement, is beyond the scope of this study.
\begin{table}[t]
 \begin{center}
  \begin{tabular}{|c||c|c|c|c|}
   \hline
   & \shortstack{Parameters\\{}}
   & \shortstack{Decay Width\\{}[GeV]}
   & \shortstack{Resonance\\[1mm] cross-section [fb]}
   & \shortstack{\\[1mm]Raw asymmetry\\around resonance} \\
   \hline \hline
   \shortstack{Case~$1^\prime$\\{}} &
   \shortstack{\\[1mm] $m_R,m_I=500,700$~GeV,\\
   $\eta_U=3, \lambda_{4,5}=1$, $\alpha_u=\pi/4$} &
   \shortstack{$S_R$ : 24.3 \\ $S_I$ : 47.7 } &
   \shortstack{$S_R$ : 60.4 \\ $S_I$ : 24.0 } &
   \shortstack{$S_R$ : $A_1=-0.127$ \\ $S_I$ : $A_1=0.103$}
   \\[1mm] \hline
   \shortstack{Case~$2^\prime$\\{}} &
   \shortstack{\\[1mm] $m_R,m_I=500,700$~GeV,\\
   $\eta_U=3, \lambda_{4,5}=8$, $\alpha_{u,4}=-\alpha_5=\pi/4$} &
   \shortstack{$S_R$ : 24.3 \\ $S_I$ : 47.7 } &
   \shortstack{$S_R$ : 43.2 \\ $S_I$ : 24.2 } &
   \shortstack{$S_R$ : $A_1=-0.122$ \\ $S_I$ : $A_1=0.129$}
   \\[1mm] \hline
   \shortstack{Case~$3^\prime$\\{}} &
   \shortstack{\\[1mm] $m_R,m_I=500,700$~GeV,\\
   $\eta_U=3, \lambda_{4,5}=1$, $\alpha_{u}=\pi/8$} &
   \shortstack{$S_R$ : 18.8 \\ $S_I$ : 52.5 } &
   \shortstack{$S_R$ : 75.8 \\ $S_I$ : 26.9 } &
   \shortstack{$S_R$ : $A_1=-0.117$ \\ $S_I$ : $A_1=0.076$}
   \\[1mm] \hline
   \shortstack{Case~$4^\prime$\\{}} &
   \shortstack{\\[1mm] $m_R=m_I=500$~GeV,\\
   $\eta_U=3, \lambda_{4,5}=1$, $\alpha_{u}=\pi/8$} &
   \shortstack{$S_R$ : 18.8 \\ $S_I$ : 29.9 } &
   \shortstack{$S_R+S_I$ : \\ \quad 118.2 } &
   \shortstack{$S_R+S_I$ : \\ \quad $A_1=-0.029$}
   \\[1mm] \hline
   \shortstack{Case~$5^\prime$\\{}} &
   \shortstack{\\[1mm] $m_R,m_I=400,1000$~GeV,\\
   $\eta_U=3, \lambda_{4,5}=1$, $\alpha_{u}=\pi/4$} &
   \shortstack{$S_R$ : 10.9 \\ $S_I$ : 79.0 } &
   \shortstack{$S_R$ : 99.2 \\ $S_I$ :  9.1 } &
   \shortstack{$S_R$ : $A_1=-0.089$ \\ $S_I$ : $A_1=0.082$}
   \\[1mm] \hline
  \end{tabular}
 \end{center}
\caption{
 Parameter values, resonance $S_{I,R}$ decay-widths and production cross-sections $\sigma(pp\to S_{I,R}\to t \bar{t} \to b \bar{b} \mu^+\mu^- \nu\bar{\nu})$ at the LHC $\sqrt{S}=14$~TeV, and  raw $CP$ asymmetry for the five cases discussed in the text with $\eta_U=3$. 
 The raw asymmetry is defined by taking into account the events with
 $|m_{\ttb}-m_{I,R}|<10$~GeV.}\label{tab1'}
\end{table}
We now have examples of large raw asymmetries in resonances that may be observable over the SM background as can seen in Figure~\ref{fig:mtt}.  

In Table~\ref{tab2'}  we examine the $CP$ asymmetries that result when all $t\bar{t}$ events are included (SM plus new resonances). In the first column we show the asymmetry diluted by SM events. We define this asymmetry using a 20~GeV $m_{t\bar{t}}$ window around the resonance.  As expected, it is roughly one order of magnitude smaller than the raw asymmetries for the resonant cross-sections shown in Table~\ref{tab1'}, and larger for $S_I$ which has a larger production cross-section. In the second column we show how the asymmetry becomes even smaller when the whole $m_{t\bar{t}}$ range is used, due to the partial cancellation between the two resonances. 
Finally, in the last column, we impose a realistic kinematic cut to simulate selecting different 
$m_{t\bar{t}}$ windows to separate the two scalar contributions. Note that the numbers in parenthesis in Table~\ref{tab2'} give the statistical
error in our simulation, which is expected to be the same order as the
statistical error for ${\cal L}=400$~[fb$^{-1}$] at the LHC. 

\begin{table}[t]
 \begin{center}
  \begin{tabular}{|c||c|c|c|c|}
   \hline
   & \shortstack{\\[1mm] Asymmetry around resonance\\including SM
   amplitudes}
   & \shortstack{Integrated asymmetry\\including SM amplitudes} &
   $M_{T}$ cut \\
   \hline \hline
   \shortstack{Case~$1^\prime$\\{}} &
   \shortstack{$S_R$ : $A_1=-0.017(2)$ \\ $S_I$ : $A_1=0.029(4)$
   } &
   $A_1=0.0002(5)$ &
   \shortstack{\\[1mm] Low : $A_1=-0.0045(10)$ \\ High :
   $A_1=0.0141(23)$ } \\[1mm]
   \hline
   \shortstack{Case~$2^\prime$\\{}} &
   \shortstack{$S_R$ : $A_1=-0.013(2)$ \\ $S_I$ : $A_1=0.021(4)$
   } &
   $A_1=-0.0007(5)$ &
   \shortstack{\\[1mm] Low : $A_1=-0.0026(10)$ \\ High : $A_1=0.0120(23)$
   } \\[1mm]
   \hline
   \shortstack{Case~$3^\prime$\\{}} &
   \shortstack{$S_R$ : $A_1=-0.026(2)$ \\ $S_I$ : $A_1=0.030(4)$
   } &
   $A_1=-0.0022(5)$ &
   \shortstack{\\[1mm] Low : $A_1=-0.0071(10)$ \\ High :
   $A_1=0.0136(23)$ } \\[1mm]
   \hline
   \shortstack{Case~$4^\prime$\\{}} &
   \shortstack{\\[1mm] $S_R+S_I$ :\\ \quad $A_1=-0.009(2)$ } &
   $A_1=-0.0003(5)$ & Low : $A_1=-0.0023(10)$  \\[1mm]
   \hline
   \shortstack{Case~$5^\prime$\\{}} &
   \shortstack{$S_R$ : $A_1=-0.021(2)$ \\ $S_I$ :
   $A_1=0.041(10)$ } &
   $A_1=-0.0016(5)$ &
   \shortstack{\\[1mm] Low : $A_1=-0.0059(8)$ \\ High : $A_1=0.023(4)$ }
   \\[1mm]
   \hline
  \end{tabular}
\caption{$CP$ asymmetries for the illustrative cases of Table~\ref{tab1'}
  including SM top-quark pair events.
  In the first column we define the asymmetry in the window
  $|m_{\ttb}-m_{I,R}|<10$~GeV.
  In the middle column, we show the integrated asymmetry over the full $m_{t\bar{t}}$ range. 
  In the last column, we illustrate how to isolate the resonances with
  more realistic cuts in the transverse mass. We use $400<M_{T}<500$~[GeV] for low mass and
  $600<M_{T}<700$~[GeV] for high mass cuts for Cases~1$^\prime$~to~4$^\prime$. 
  For Case~5$^\prime$ we use $300<M_{T}<400$~[GeV] for low mass and
  $800<M_{T}<1000$~[GeV] for high mass cuts. 
  The numbers in parenthesis are the statistical error in our simulation,
  which is expected to be the same order as the statistical error for
  ${\cal L}=400$~[fb$^{-1}$] at the LHC.
  }\label{tab2'}
 \end{center}
\end{table}

\section{Results and Conclusion}

We have investigated the $CP$ violating asymmetries that result in
$t\bar{t}$ events at the LHC from new sources of $CP$ violation
associated with color-octet scalars.
We have considered the effect of two new sources of $CP$  violation: a
phase in the couplings of the new resonances $S_{I,R}$ to top-quarks;
and two phases in the quartic couplings of the scalar potential. We find
that the former is responsible for much larger $CP$ violating effects
than the latter.
We have shown that these models typically induce large raw asymmetries
that can reach 12\%.

Observation of the asymmetries is contingent to observation of the new
resonance itself and we have presented a rough numerical simulation that
illustrates this.
An optimistic sensitivity for the asymmetry {\it using only
the dimuon channel} could be $\delta{A}_1\sim 3\times 10^{-3}$ at the
LHC with an integrated luminosity of ${\cal L}=10$~[fb$^{-1}$], when the
events in full kinematical region are taken into account.
In Table~\ref{tab2'} we have shown several examples utilizing
kinematical cuts with resulting asymmetries enhanced as large as a few
percent by enhancing the scalar resoance contribution and avoiding
cancellation between the two scalar contributions, which could be
observed at the LHC.

We have presented our analysis for the dimuon channel as this is the cleanest one. However, our study can be easily extended to other top-quark decay channels to increase statistics. We have studied the $CP$ asymmetries only in the top-quark pair production channel via one new resonance. Other channels, such as $SS$ pair production, also exhibit $CP$ violating asymmetries and may be preferable in scenarios where the $SS$ production cross-section exceeds the single $S$ production cross-section.

\begin{acknowledgments}
This work was partially supported by NSC, NCTS, SJTU 985 grant, and
Excellent Research Projects of
National Taiwan University (NTU-98R0526), and in part by DOE under contract number DE-FG02-01ER41155.
G.V. thanks the National Taiwan University, Taipei, Taiwan for their hospitality.

\end{acknowledgments}

\appendix

\section{CP Violation with a Higgs boson}

We review the salient features of $CP$ violation in $t\bar{t}$ production  at the LHC induced by a new Higgs-boson as discussed in Ref.~\cite{cphiggs}. The $CP$ violation in a suitable extended Higgs sector manifests itself in the form of a neutral Higgs mass eigenstate that has both scalar and pseudo-scalar couplings to the top-quark. In general these couplings can be written as
\begin{eqnarray}
{\cal L}&=&-\frac{m_t}{v}H\bar{t} (A+iB\gamma_5)t,
\label{tlag}
\end{eqnarray}
where $A,B$ are real and $A=1,~B=0$ corresponds to the standard model with one Higgs doublet. Multi-Higgs models achieve maximal $CP$ violation when $A=B$, and these couplings reach the Weinberg unitarity bound, $|AB| \lsim \frac{1}{\sqrt{2}}$~\cite{Weinberg:1990me}.

This Higgs-boson is produced at the LHC mostly via gluon fusion. Both the scalar and pseudoscalar cases have been considered in the literature before and these results at leading order can be summarized by the effective couplings
\begin{eqnarray}
{\cal L}&=&\left[F_a G_{\mu\nu}G^{\mu\nu}+F_b \tilde{G}_{\mu\nu}G^{\mu\nu}
\right] H
\end{eqnarray}
The two form factors $F_a(s)$ and $F_b(s)$ can be found, for example in Ref.~\cite{Spira:1995rr}. For the kinematic regime in which the Higgs boson is heavier that a $t\bar{t}$ pair they are given by
\begin{eqnarray}
F_a &=& \left(\sqrt{2} G_F \right)^{1/2} A \frac{\alpha_s}{12\pi} \, 3I_q(z) \nonumber \\
F_b &=& -\left(\sqrt{2} G_F \right)^{1/2} B \frac{\alpha_s}{8\pi} z f(z)
\end{eqnarray}
with $z = m_t^2/m_H^2$, the functions $I_q(z)$ and $f(z)$ arise from the one-loop contribution of a top-quark loop and were given in Eq.~(\ref{loopintegrals}).

The origin of the correlation, Eq.~(\ref{cpcorrelation}) in this model is a $CP$ violating term in the invariant matrix element squared for $gg\to t\bar{t} \to b\bar{b}\mu^+\mu^-\nu\bar\nu$ of the form
\begin{eqnarray}
\left| {\cal M}\right|^2 &=&  C_1(s,t,u) \ \epsilon(p_t,p_{\bar{t}},p_{\mu^+},p_{\mu^-}) + \cdots
\label{tripprod}
\end{eqnarray}
The function $C_1(s,t,u)$ can be written in a compact form reflecting two contributions: the $s$ channel Higgs amplitude squared; and  the interference between the $t$ and $u$ channels ($gg \to t\bar{t}$ in a color singlet state) and the $s$-channel Higgs amplitude. They are given by
\begin{eqnarray}
C_1(s,t,u) &=& 24 \,  K_{\ell\ell}\, AB(|F_a|^2+|F_b|^2)
\frac{s^2}{(s-m_H^2)^2+m_H^2\Gamma_H^2} \frac{m_t^4}{v^2}\nonumber \\
&+& 8 g_s^2 \,  K_{\ell\ell}\, \left(B {\rm Re} (F_a) + A {\rm Re} (F_b)\right)\frac{s^2(s-m_H^2)}{((s-m_H^2)^2+m_H^2\Gamma_H^2)(s^2-(t-u)^2)} \frac{m_t^4}{v}. \nonumber \\
K_{\ell\ell} &\equiv & g^8\,  \left(p_b\cdot p_\nu\right)\left( p_{\bar{b}}\cdot p_{\bar{\nu}} \right)\, \left(\frac{\pi}{m_t\Gamma_t}\right)^2\left(\frac{\pi}{M_W\Gamma_W}\right)^2 \nonumber \\
&\times&  \delta(p_t^2-m_t^2)\delta(p_{\bar{t}}^2-m_t^2)
\delta(p_{W^+}^2-M_W^2)  \delta(p_{W^-}^2-M_W^2).
\label{s2singlet}
\end{eqnarray}
The delta functions in this expression reflect the use of the narrow-width approximation for all top-quark and $W$-boson propagators. If the Higgs boson is also narrow, the expression simplifies further to
\begin{eqnarray}
C_1(s,t,u) &=& 24 \,  K_{\ell\ell}\,AB(|F_a|^2+|F_b|^2)\frac{s^2 m_t^4}{v^2}\left(\frac{\pi}{m_H\Gamma_H}\right)\ \delta(s-m_H^2)
\label{s2nw}
\end{eqnarray}

\section{CP Violation with color-octet scalars}

In the color-octet model discussed in this paper, the $CP$ violation in the process $gg\to t\bar{t} \to b\bar{b}\mu^+\mu^-\nu\bar\nu$ takes the same form as Eq.~(\ref{tripprod}) with the form factor of Eq.~(\ref{s2nw}) replaced by the sum of the contributions from the octet neutral scalars $S_R$ and $S_I$. The overall color factor changes from $24$ to $20/3$, and in the narrow width approximation we have
\begin{eqnarray}
C_1(s,t,u) &=&  \frac{20 K_{\ell\ell}}{3} \,   s^2 m_t^2 \left(a_R b_R(|F^a_R|^2+|F^b_R|^2)\left(\frac{\pi}{m_R\Gamma_R}\right)\ \delta(s-m_R^2) \right. \nonumber \\
&+& \left. a_Ib_I(|F^a_I|^2+|F^b_I|^2)\left(\frac{\pi}{m_I\Gamma_I}\right)\ \delta(s-m_I^2)\right) \nonumber \\
&=& \frac{20 K_{\ell\ell}}{3} \, \frac{s^2 m_t^4}{v^2}\frac{\eta_U^2s_u c_u}{16}
\left( \left(\frac{\pi D_R}{m_R\Gamma_R}\right)\ \delta(s-m_R^2) +
 \left(\frac{\pi D_I}{m_I\Gamma_I}\right)\
 \delta(s-m_I^2)\right).
\label{s2octet}
\end{eqnarray}
Where we have introduced  for compactness the notation, $s_i \equiv \sin\alpha_i,\ c_i \equiv \cos\alpha_i$. The coefficients $D_{I,R}$ can be written as
\begin{eqnarray}
D_I&=& 4\eta_U^2 
 \left\{c^2_u+s^2_u\left(1-\frac{4m_t^2}{m_I^2}\right)^2\right\}
 \frac{m_t^4}{m_I^4}\left|f\left(\frac{m_t^2}{m_I^2}\right)\right|^2
 \nonumber \\
 &-& 72\eta_U s_u\left(\lambda_4 s_4+\lambda_5 s_5\right)
  {\rm Re}\left\{\left[\frac{5}{6}I_s(1)+\frac{1}{6}I_s\left(\frac{m_R^2}{m_I^2}\right)\right]
  I_q^\star\left(\frac{m_t^2}{m_I^2}\right)\right\} \nonumber \\
 &+& 81\left(\lambda_4 s_4+\lambda_5 s_5\right)^2
  \left|\frac{5}{6}I_s(1)+\frac{1}{6}I_s\left(\frac{m_R^2}{m_I^2}\right)\right|^2 \\
 D_R&=& -4\eta_U^2 
  \left\{ s^2_u+c^2_u\left(1-\frac{4m_t^2}{m_R^2}\right)^2 \right\}
  \frac{m_t^4}{m_R^4} \left|f\left(\frac{m_t^2}{m_R^2}\right)\right|^2
  \nonumber \\
 &+& 72\eta_U c_u\left(\lambda_4 c_4+\lambda_5 c_5\right)
  {\rm Re}\left\{\left[\frac{1}{2}I_s(1)+\frac{1}{2}I_s\left(\frac{m_I^2}{m_R^2}\right)\right]
  I_q^\star\left(\frac{m_t^2}{m_R^2}\right)\right\} \nonumber \\
 &-& 81\left(\lambda_4 c_4+\lambda_5 c_5\right)^2
  \left|\frac{1}{2}I_s(1)+\frac{1}{2}I_s\left(\frac{m_I^2}{m_R^2}\right)\right|^2 
 \label{dianddr}
\end{eqnarray}

If the two neutral scalars have masses that are close to each other (recall that $m_R^2 - m_I^2 = \lambda_3 v^2$) then the form factor becomes
\begin{eqnarray}
C_1(s,t,u) &=&  \frac{20 K_{\ell\ell}}{3} \,   s^2 m_t^2 \left[
 \frac{a_R b_R(|F^a_R|^2+|F^b_R|^2)}{(s-m_R^2)^2+m_R^2\Gamma_R^2} +\frac{a_Ib_I(|F^a_I|^2+|F^b_I|^2)}{(s-m_I^2)^2+m_I^2\Gamma_I^2} \right. \nonumber \\
&+&(s-m_R^2)(s-m_I^2)\left. \frac{ {\rm Re}\left(a_Rb_I(F^a_RF^{a\star}_I +F^b_RF^{b\star}_I)+a_Ib_R(F^{a\star}_RF^{a}_I +F^{b\star}_RF^{b}_I)\right)}{2((s-m_I^2)^2+m_I^2\Gamma_I^2)((s-m_R^2)^2+m_R^2\Gamma_R^2)} \right] \nonumber \\
&+& \frac{20 g_s^2   K_{\ell\ell}}{3} \frac{s^2m_t^4}{v(s^2-(t-u)^2)}
 \left[\frac{\left(b_R {\rm Re} (F^a_R) + a_R{\rm Re} (F^b_R)\right)(s-m_R^2)}{((s-m_R^2)^2+m_R^2\Gamma_R^2)} \right.\nonumber \\
&+& \left.\frac{\left(b_I {\rm Re} (F^a_I) + a_I{\rm Re} (F^b_I)\right)(s-m_I^2)}{((s-m_I^2)^2+m_I^2\Gamma_I^2)}\right] \nonumber \\
&=& \frac{20 K_{\ell\ell}}{3} \,   \frac{s^2 m_t^4}{16v^2}  \frac{\eta_U^2}{(s-m_S^2)^2+m_S^2\Gamma_S^2} 9\left(\lambda_4\sin(\alpha_4-\alpha_u)+\lambda_5\sin(\alpha_5-\alpha_u) \right) \nonumber \\
&\times & \left( -4 \eta_U I_s(1) {\rm Re}(I_q(m_t^2/m_S^2))   + 9I_s^2(1)  \left(\lambda_4\cos(\alpha_4-\alpha_u)+\lambda_5\cos(\alpha_5-\alpha_u) \right) \right) \nonumber \\
&+& {\cal O}(m_R-m_I)
 \label{ffnearres}
\end{eqnarray}
where in the last line we have used $m_S\simeq m_R\simeq m_I$ and similarly for $\Gamma_S$.
The first two terms correspond to those in Eq.~(\ref{s2octet}) without
the narrow width approximation for $S_{I,R}$, and the third term arises
from the interference between the two resonances. Next, we have also
kept the term arising from the interference of the scalar exchange
amplitude and the QCD background, analogous to the second term in
Eq.~(\ref{s2singlet}). For the color-octet case, the overall color
factor in this term changes from $4$ to $10/3$ and the interference
still occurs only with the $t$ and $u$ channels of the QCD
amplitude. The $s$-channel gluon exchange diagram has a color structure
that does not interfere with the color-octet scalar exchange
amplitudes. As can be seen from the second expression, the two
resonances tend to cancel each other out. For degenerate resonances only
a small term proportional to the scalar loops remains. The corrections
to this limit, proportional to $(m_R-m_I)$ are too cumbersome to write
out explicitly and are best studied numerically.

\section{Decay width of $S_{I,R}$}

Dominant decay mode of $S_{I,R}$ is $t\bar{t}$ mode.
The partial decay width is given as follows~\cite{Manohar:2006ga};
\begin{align}
 &\Gamma(S_R\to t\bar{t}) = 
 \frac{m_R\eta_U^2}{16\pi}\frac{m_t^2}{v^2}
 \sqrt{1-\frac{4m_t^2}{m_R^2}}
 \left[1-\frac{4m_t^2}{m_R^2}c_u^2\right], \\
 &\Gamma(S_I\to t\bar{t}) = 
 \frac{m_I\eta_U^2}{16\pi}\frac{m_t^2}{v^2}
 \sqrt{1-\frac{4m_t^2}{m_I^2}}
 \left[1-\frac{4m_t^2}{m_I^2}s_u^2\right].
\end{align}

The partial decay width into $gg$, which is related to the production
cross-section, is give as follows;
\begin{align}
 \Gamma(S_R\to gg) &=
 \frac{G_F m_R^3C_1\alpha_S^2}{\sqrt{2}\,2^{10}\,\pi^3}
 \bigg[ \eta_U^2 c_u^2 \left|I_q\left(\frac{m_t^2}{m_R^2}\right)\right|^2
 \nonumber \\
 & -\frac{9}{2}\frac{v^2}{m_R^2}\eta_U
 c_u (\lambda_4 c_4+\lambda_5 c_5)\, {\rm Re}\left\{
 \left[\frac{1}{2}I_s\left(1\right)
 +\frac{1}{2}I_s\left(\frac{m_I^2}{m_R^2}\right)\right]
 I_q^\star\left(\frac{m_t^2}{m_R^2}\right)\right\}
 \nonumber \\
 & + \frac{81}{16}\frac{v^4}{m_R^4}
 (\lambda_4 c_4+\lambda_5 c_5)^2
 \left|\frac{1}{2}I_s\left(1\right)
 +\frac{1}{2}I_s\left(\frac{m_I^2}{m_R^2}\right)\right|^2
 + \eta_U^2 s_u^2 \frac{m_t^4}{m_R^4}
 \left|f\left(\frac{m_t^2}{m_R^2}\right)\right|^2 \bigg],
 \label{SRgg} \\[4mm]
 \Gamma(S_I\to gg) &=
 \frac{G_F m_I^3C_1\alpha_S^2}{\sqrt{2}\,2^{10}\,\pi^3}
 \bigg[ \eta_U^2 s_u^2 \left|I_q\left(\frac{m_t^2}{m_I^2}\right)\right|^2
 \nonumber \\
 & -\frac{9}{2}\frac{v^2}{m_I^2}\eta_U
 s_u (\lambda_4 s_4+\lambda_5 s_5) \, {\rm Re}\left\{
 \left[\frac{5}{6}I_s\left(1\right)
 +\frac{1}{6}I_s\left(\frac{m_R^2}{m_I^2}\right)\right]
 I_q^\star\left(\frac{m_t^2}{m_I^2}\right)\right\}
 \nonumber \\
 & + \frac{81}{16}\frac{v^4}{m_I^4}
 (\lambda_4 s_4+\lambda_5 s_5)^2
 \left|\frac{5}{6}I_s\left(1\right)
 +\frac{1}{6}I_s\left(\frac{m_R^2}{m_I^2}\right)\right|^2
 + \eta_U^2 c_u^2 \frac{m_t^4}{m_I^4}
 \left|f\left(\frac{m_t^2}{m_I^2}\right)\right|^2 \bigg],
 \label{SIgg}
\end{align}
where a color-factor $C_1=40/3$.
In the $\alpha_{u,4,5}\to 0$ limit, Eq.~(\ref{SRgg}) and (\ref{SIgg})
coincide with Eq.~(25) and (29) of Ref.~\cite{Gresham:2007ri},
respectively.


\begin{thebibliography}{99}

\bibitem{tprods}
%\cite{Donoghue:1987ax}
%\bibitem{Donoghue:1987ax}
  J.~F.~Donoghue and G.~Valencia,
  %``Searching For CP Violation In Jet Physics,''
  Phys.\ Rev.\ Lett.\  {\bf 58}, 451 (1987)
  [Erratum-ibid.\  {\bf 60}, 243 (1988)];
  %%CITATION = PRLTA,58,451;%%
%\cite{Gavela:1988jx}
%\bibitem{Gavela:1988jx}
  M.~B.~Gavela, F.~Iddir, A.~Le Yaouanc, L.~Oliver, O.~Pene and J.~C.~Raynal,
  %``T ODD AND CP ODD APLANARITIES IN e+ e- COLLIDERS,''
  Phys.\ Rev.\  D {\bf 39}, 1870 (1989);
  %%CITATION = PHRVA,D39,1870;%%
%\cite{Kamionkowski:1989vj}
%\bibitem{Kamionkowski:1989vj}
  M.~P.~Kamionkowski,
  %``SEARCHING FOR CP VIOLATION IN 'CHARGE BLIND' JETS,''
  Phys.\ Rev.\  D {\bf 41}, 1672 (1990).
  %%CITATION = PHRVA,D41,1672;%%

\bibitem{cplhc}
%\cite{Antipin:2008zx}
%\bibitem{Antipin:2008zx}
  O.~Antipin, G.~Valencia,
  %``T-odd correlations from CP violating anomalous top-quark couplings revisited,''
  Phys.\ Rev.\  {\bf D79 } (2009)  013013.
  [arXiv:0807.1295 [hep-ph]];
  %\cite{Gupta:2009wu}
%\bibitem{Gupta:2009wu}
  S.~K.~Gupta, A.~S.~Mete, G.~Valencia,
  %``CP violating anomalous top-quark couplings at the LHC,''
  Phys.\ Rev.\  {\bf D80 } (2009)  034013.
  [arXiv:0905.1074 [hep-ph]];
%\cite{Gupta:2009eq}
%\bibitem{Gupta:2009eq}
  S.~K.~Gupta, G.~Valencia,
  %``CP-odd correlations using jet momenta from tt_bar events at the Tevatron,''
  Phys.\ Rev.\  {\bf D81 } (2010)  034013.
  [arXiv:0912.0707 [hep-ph]].

\bibitem{cphiggs}
  %\cite{Chang:1992tu}
%\bibitem{Chang:1992tu}
  D.~Chang and W.~Y.~Keung,
  %``CP violation in the decay of neutral Higgs boson into t - anti-t and W+- W-,''
  Phys.\ Lett.\ B {\bf 305}, 261 (1993)
  [arXiv:hep-ph/9301265]; 
 %\cite{Gunion:1996xu}
%\bibitem{Gunion:1996xu}
  J.~F.~Gunion, X.~-G.~He,
  %``Determining the CP nature of a neutral Higgs boson at the LHC,''
  Phys.\ Rev.\ Lett.\  {\bf 76}, 4468-4471 (1996).
  [hep-ph/9602226]; 
  %\cite{Bernreuther:1993df}
%\bibitem{Bernreuther:1993df}
  W.~Bernreuther and A.~Brandenburg,
  %``Signatures of Higgs sector CP violation in top quark pair production at
  %proton proton supercolliders,''
  Phys.\ Lett.\ B {\bf 314}, 104 (1993); 
  %%CITATION = PHLTA,B314,104;%%   
%\cite{Bernreuther:1993hq}
%\bibitem{Bernreuther:1993hq}
  W.~Bernreuther and A.~Brandenburg,
  %``Tracing CP violation in the production of top quark pairs by multiple TeV
  %proton proton collisions,''
  Phys.\ Rev.\ D {\bf 49}, 4481 (1994)
  [arXiv:hep-ph/9312210];
  %%CITATION = HEP-PH 9312210;%%
 %\cite{Bernreuther:1998qv}
%\bibitem{Bernreuther:1998qv}
  W.~Bernreuther, A.~Brandenburg and M.~Flesch,
  %``Effects of Higgs sector CP violation in top-quark pair production at  the
  %LHC,''
  arXiv:hep-ph/9812387;
  %%CITATION = HEP-PH 9812387;%%
%\cite{Valencia:2005cx}
%\bibitem{Valencia:2005cx}
  G.~Valencia, Y.~Wang,
  %``New CP-odd observable in H ---> t anti-t,''
  Phys.\ Rev.\  {\bf D73 } (2006)  053009.
  [hep-ph/0512127].

\bibitem{others}
  %\cite{Berger:2011wh}
%\bibitem{Berger:2011wh}
  J.~Berger, M.~Blanke, Y.~Grossman,
  %``A new CP violating observable for the LHC,''
  JHEP {\bf 1108}, 033 (2011).
  [arXiv:1105.0672 [hep-ph]];
%\cite{Christensen:2010pf}
%\bibitem{Christensen:2010pf}
  N.~D.~Christensen, T.~Han, Y.~Li,
  %``Testing CP Violation in ZZH Interactions at the LHC,''
  Phys.\ Lett.\  {\bf B693}, 28-35 (2010).
  [arXiv:1005.5393 [hep-ph]];
   %\cite{MoortgatPick:2009jy}
%\bibitem{MoortgatPick:2009jy}
  G.~Moortgat-Pick, K.~Rolbiecki, J.~Tattersall, P.~Wienemann,
  %``Probing CP Violation with and without Momentum Reconstruction at the LHC,''
  JHEP {\bf 1001}, 004 (2010).
  [arXiv:0908.2631 [hep-ph]];
   %\cite{Godbole:2007cn}
%\bibitem{Godbole:2007cn}
  R.~M.~Godbole, D.~J.~Miller, 2, M.~M.~Muhlleitner,
  %``Aspects of CP violation in the H ZZ coupling at the LHC,''
  JHEP {\bf 0712}, 031 (2007).
  [arXiv:0708.0458 [hep-ph]].


%\cite{Hagiwara:2007sz}
\bibitem{Hagiwara:2007sz}
  K.~Hagiwara, K.~Mawatari and H.~Yokoya,
  %``T-odd asymmetries in radiative top-quark decays,''
  JHEP {\bf 0712}, 041 (2007)
  [arXiv:0707.3194 [hep-ph]].
  %%CITATION = JHEPA,0712,041;%%


%\cite{Manohar:2006ga}
\bibitem{Manohar:2006ga}
  A.~V.~Manohar, M.~B.~Wise,
  %``Flavor changing neutral currents, an extended scalar sector, and the Higgs production rate at the CERN LHC,''
  Phys.\ Rev.\  {\bf D74}, 035009 (2006).
  [hep-ph/0606172].

%\cite{Gresham:2007ri}
\bibitem{Gresham:2007ri}
  M.~I.~Gresham, M.~B.~Wise,
  %``Color octet scalar production at the LHC,''
  Phys.\ Rev.\  {\bf D76}, 075003 (2007).
  [arXiv:0706.0909 [hep-ph]].

\bibitem{noreson}
%\cite{Aad:2011fq}
%\bibitem{Aad:2011fq}
  G.~Aad {\it et al.} [ ATLAS Collaboration ],
  %``Search for New Physics in the Dijet Mass Distribution using 1 fb^-1 of pp Collision Data at sqrt(s) = 7 TeV collected by the ATLAS Detector,''
   [arXiv:1108.6311 [hep-ex]]; 
 %\cite{Chatrchyan:2011wq}
%\bibitem{Chatrchyan:2011wq}
  S.~Chatrchyan {\it et al.} [ CMS Collaboration ],
  %``Search for Resonances in the Dijet Mass spectrum from  $7$ TeV $pp$ Collisions at CMS,''
Phys. Lett. {\bf B704}, 123 (2011).  


%\cite{Burgess:2009wm}
\bibitem{Burgess:2009wm}
  C.~P.~Burgess, M.~Trott and S.~Zuberi,
  %``Light Octet Scalars, a Heavy Higgs and Minimal Flavour Violation,''
  JHEP {\bf 0909}, 082 (2009)
  [arXiv:0907.2696 [hep-ph]].
  %%CITATION = JHEPA,0909,082;%%

%\cite{Carpenter:2011yj}
\bibitem{Carpenter:2011yj}
  L.~M.~Carpenter and S.~Mantry,
  %``Color-Octet, Electroweak-Doublet Scalars and the CDF Dijet Anomaly,''
  arXiv:1104.5528 [hep-ph].
  %%CITATION = ARXIV:1104.5528;%%


\bibitem{madgraph}
%\cite{Stelzer:1994ta}
%\bibitem{Stelzer:1994ta}
  T.~Stelzer and W.~F.~Long,
  %``Automatic generation of tree level helicity amplitudes,''
  Comput.\ Phys.\ Commun.\  {\bf 81}, 357 (1994)
  [arXiv:hep-ph/9401258];
  %%CITATION = CPHCB,81,357;%%
%\cite{Alwall:2007st}
%\bibitem{Alwall:2007st}
  J.~Alwall {\it et al.},
  %``MadGraph/MadEvent v4: The New Web Generation,''
  JHEP {\bf 0709}, 028 (2007)
  [arXiv:0706.2334 [hep-ph]];
  %%CITATION = JHEPA,0709,028;%%
  %\cite{Alwall:2011uj}
%\bibitem{Alwall:2011uj}
  J.~Alwall, M.~Herquet, F.~Maltoni, O.~Mattelaer, T.~Stelzer,
  %``MadGraph 5 : Going Beyond,''
  JHEP {\bf 1106}, 128 (2011).
  [arXiv:1106.0522 [hep-ph]].

%\cite{Pumplin:2002vw}
\bibitem{Pumplin:2002vw}
  J.~Pumplin, D.~R.~Stump, J.~Huston, H.~L.~Lai, P.~M.~Nadolsky and W.~K.~Tung,
  %``New generation of parton distributions with uncertainties from global QCD
  %analysis,''
  JHEP {\bf 0207}, 012 (2002)
  [arXiv:hep-ph/0201195].
  %%CITATION = JHEPA,0207,012;%%

%\cite{Weinberg:1990me}
\bibitem{Weinberg:1990me}
  S.~Weinberg,
  %``Unitarity Constraints On Cp Nonconservation In Higgs Exchange,''
  Phys.\ Rev.\  {\bf D42}, 860-866 (1990).

%\cite{Spira:1995rr}
\bibitem{Spira:1995rr}
  M.~Spira, A.~Djouadi, D.~Graudenz, P.~M.~Zerwas,
  %``Higgs boson production at the LHC,''
  Nucl.\ Phys.\  {\bf B453}, 17-82 (1995).
  [arXiv:hep-ph/9504378 [hep-ph]].



%%\cite{Enkhbat:2011qz}
%\bibitem{Enkhbat:2011qz}
%  T.~Enkhbat, X.~G.~He, Y.~Mimura and H.~Yokoya,
%  %``Colored Scalars And The CDF $W+$dijet Excess,''
%  arXiv:1105.2699 [hep-ph].
%  %%CITATION = ARXIV:1105.2699;%%
%%\cite{He:2011ti}
%\bibitem{He:2011ti}
%  X.~G.~He and G.~Valencia,
%  %``An extended scalar sector to address the tension between a fourth
%  %generation and Higgs searches at the LHC,''
%  arXiv:1108.0222 [hep-ph].
%  %%CITATION = ARXIV:1108.0222;%%



\end{thebibliography}
\end{document}